\documentstyle[aps,graphicx,epsf,floats]{revtex}

\begin{document}

\draft

\twocolumn[\hsize\textwidth\columnwidth\hsize\csname@twocolumnfalse%
\endcsname

\title{Specific-Heat Exponent of Random-Field Systems via
Ground-State Calculations 
}

\author{A.K. Hartmann\cite{alex:email}
and 
A.P. Young\cite{peter:email}
}
\address{Department of Physics, 
University of California, Santa Cruz CA 95064, USA}

\date{\today}
\maketitle
\begin{abstract}
Exact ground states of three-dimensional 
random field Ising magnets (RFIM) with Gaussian distribution of the
disorder are calculated
using graph-theoretical algorithms. Systems for different strengths $h$
of the random fields and sizes up to $N=96^3$ are considered.
By numerically differentiating the bond-energy with respect to $h$
a specific-heat like quantity is obtained, which does not appear to diverge at
the critical point but rather exhibits a cusp. We also consider the effect of a
small
uniform magnetic field, which allows us to calculate the $T=0$
susceptibility. From a finite-size scaling analysis, we obtain the
critical exponents $\nu=1.32(7)$, $\alpha=-0.63(7)$, $\eta=0.50(3)$ and find
that the
critical strength of the random field is $h_c=2.28(1)$. We discuss the
significance of the result that $\alpha$ appears to be strongly negative.

\end{abstract}

]

\section{Introduction}
\label{sec:intro}

The random field Ising model\cite{im75} has been extensively
stud\-ied\cite{by1991,rieger1995,young1999} both because of its
interest as a ``simple'' frustrated system and because of its relevance to
experiments, especially those on the diluted antiferromagnet in a uniform
field\cite{fa79}.
The RFIM Hamiltonian is given by

\begin{equation}
{\cal H}=-J\sum_{\langle i,j \rangle} S_i S_j - \sum_i h_i S_i ,
\label{eq:ham}
\end{equation}
where the $S_i = \pm 1$ are Ising spins, $J$ is the interaction energy
between nearest neighbors, and $h_i$ is the random field.
The values $h_i$ are independently distributed according a
Gaussian distribution with mean 0 and standard deviation $h$,
i.e. the probability distribution is

\begin{equation}
P(h_i)=\frac{1}{\sqrt{2\pi}\, h}\exp\left(-{h_i^2 \over 2 h^2} \right) .
\end{equation}
We shall consider three-dimensional lattices with periodic boundary
condition and $N=L^3$ spins.

\begin{figure}
\begin{center}
\epsfxsize=\columnwidth
\epsfbox{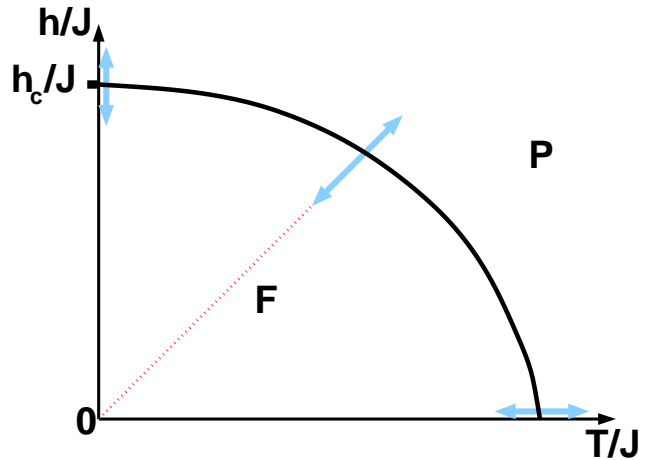}
\end{center}
\caption{
A sketch of the phase boundary of the random field Ising model. The
ferromagnetic phase is denoted by ``F'' and the paramagnetic phase by ``P''.
The critical value of the random field at $T=0$ is denoted by $h_c$. The lines
with arrows at both ends indicate the path followed by varying $J$ for some
fixed value of $h$ and $T$.
}
\label{phase}
\end{figure}

A sketch of the phase boundary is shown in Fig.~\ref{phase}. At low values of
the random field and temperature $T$, the system is in a ferromagnetic phase,
and at high temperatures or random fields, the system is paramagnetic.

In this paper we shall be interested in the values of the critical exponents
along the phase boundary. The random field is a relevant perturbation at the
pure (i.e. $h=0$) fixed point, and the random-field
fixed point is at $T=0$ \cite{bm85,fisher85}. Hence, the
critical behavior is the same everywhere along the phase boundary
in Fig.~\ref{phase} (assuming
that the transition is always second order) except for $h = 0$. 
We can therefore determine the critical behavior by staying at $T=0$ and
crossing the phase boundary at $h=h_c$, see Fig.~\ref{phase},
which is convenient, because we can
determine the ground states of large lattices exactly using efficient
optimization
algorithms\cite{angles-d-auriac1997b,rieger1998,alava2001,opt-phys2001},
as discussed in Sec.~\ref{sec:numerics}. This has the
advantage that one can
study much larger systems than it is possible in Monte Carlo simulations, and,
for each realization,
there are no statistical errors or equilibration problems.

Using these ground state
techniques, most of the critical exponents have been determined
with some precision; for a thorough recent study see Ref.~\onlinecite{mf01}.
Most of these exponents are consistent with scaling relations. However, as we
shall discuss in Sec.~\ref{sec:summary},
those
scaling relations predict a specific-heat exponent $\alpha$ close to
zero, while Monte Carlo data on fairly small sizes\cite{rieger1995b}
($L \le 16$) find $\alpha/\nu = -0.45 \pm 0.05$, where $\nu$ is the correlation
length exponent (which has a value slightly greater than unity, as discussed
in Secs.~\ref{sec:results} and \ref{sec:summary}).
Interestingly,
experiments find \cite{belanger} a logarithmic
divergence, corresponding to a specific heat exponent $\alpha
= 0$, as expected from scaling.

In order to try to resolve this puzzle, we calculate here the specific
heat exponent for the RFIM using 
{\em much}\/
larger sizes ($L \le 96$) than in the Monte Carlo work\cite{rieger1995b}, 
by using optimization methods to determine exact ground states.
We also find a strongly negative value for $\alpha$,
$\alpha/\nu = -0.48 \pm 0.05$, consistent with the earlier Monte Carlo
data\cite{rieger1995b}, but in disagreement with ex\-pe\-ri\-ment and
apparently in violation of scaling. In Sec.~\ref{sec:summary} we will
discuss possible ways round this discrepancy. In addition, we determine the
susceptibility, which, to our knowledge, has not been directly computed before
using ground-state methods. Our results are consistent with earlier
calculations.

\section{Numerical techniques}

\label{sec:numerics}
We used well known algorithms 
\cite{angles-d-auriac1997b,rieger1998,alava2001,opt-phys2001}
from graph theory
\cite{swamy,claibo,knoedel} to calculate the ground state of a system
at given random-field strength $h$. To implement them we
applied some algorithms from the LEDA library \cite{leda1999}.
The calculation works by transforming the
system into a network \cite{picard1}, and calculating the maximum flow
in polynomial time 
\cite{traeff,tarjan,goldberg1988,cherkassky1997,goldberg1998}. 
The first results of applying these algorithms to random-field
systems can be found in Ref. \onlinecite{ogielski}. In Ref.
\onlinecite{alex-rfim3} these methods were applied to obtain the exponents 
for the magnetization, the disconnected susceptibility and the
correlation length from ground-state calculations up to size $L=80$.
Other exact ground-state calculation of the RFIM can be found
in Refs. \onlinecite{angles-d-auriac1997,sourlas1999,nowak1998,mf01}.
Note that in cases where the ground-state
is degenerate \cite{foot-degenerate}
it is possible to calculate all the ground-states in one
sweep \cite{picard2}, see also Refs. \cite{alex-daff2,bastea1998}. For the
RFIM with a Gaussian distribution of fields, the ground state is
non-degenerate, except for a 
two-fold degeneracy at certain values of the randomness, where the
ground state changes, see
Sec.~\ref{sec:quantities},
so it is sufficient to calculate just one ground state.

\section{Quantities of Interest}
\label{sec:quantities}
In zero random field, the
specific-heat exponent is obtained from the singularity in the
second derivative of the free energy
with respect to temperature. More generally it is determined from
the singularity obtained by varying a parameter which crosses
the phase boundary from the paramagnetic phase to the ferromagnetic phase.
From Fig.~\ref{phase} we see that this can be conveniently accomplished by
keeping the ratio of $h/J$ to $T/J$ fixed, i.e. by varying $J$. The first
derivative of the free energy (per spin) $F$
with respect to $J$, which we call the ``bond energy''
$E_J$, is
given by
\begin{equation}
E_J \equiv {\partial F \over \partial J}
= -{1 \over N} \sum_{\langle i,j\rangle} \langle S_i S_j \rangle ,
\label{eq:EJ}
\end{equation}
where $\langle \cdots \rangle$ is a thermal average, and
the sum is over nearest-neighbor pairs. $E_J$ has an energy-like
singularity in the vicinity of the phase boundary. For $h=0$ it is
{\em precisely}\/ the energy, apart from an overall factor of $J$.

The total energy per spin, $E$, is given by
\begin{equation}
E = J E_J + h E_h ,
\label{totalE}
\end{equation}
where  the ``field energy'' $E_h$ is given by
\begin{equation}
E_h \equiv {\partial F \over \partial h}
= -{1 \over N} \sum_i \left( {h_i \over h} \right) \langle S_i \rangle .
\end{equation}

\begin{figure}[htb]
\begin{center}
\epsfxsize=\columnwidth
\epsfbox{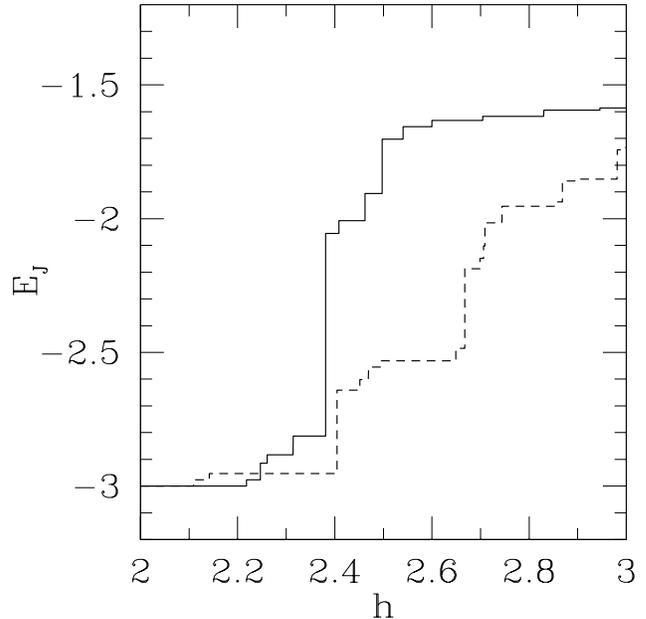}
\end{center}
\caption{Bond energy per spin, $E_J$, defined in Eq.~(\ref{eq:EJ}),
for two $L=8$ samples as a function of the 
random-field strength $h$.}
\label{figSampleJumps}
\end{figure}

Having differentiated {\em analytically}
with respect to $J$, we now set $J=1$, consider $T=0$
only, and obtain a specific heat-like quantity by
differentiating $E_J$ {\em numerically}\/ with respect to the random field $h$. 
We emphasize that it is not necessary to vary the temperature in order to
observe the specific heat singularity. To observe this singularity
the direction in which the phase boundary is crossed 
must have a projection on to the correct scaling field, which
means that the phase boundary should not be approached
tangentially. 
The angle at which the phase boundary is approached will affect the size of
{\em corrections} to scaling by mixing in a varying amount of irrelevant
operators, but the asymptotic behavior will always be the same (as long as the
approach is not tangential).

To avoid confusion we point out that the role taken by the
free energy at finite-$T$ is played by the energy at $T=0$, since the two are
equal in this limit.  More precisely,
the {\em energy}\/ singularity at $T=0$ has the form
$\epsilon^{2-\alpha}$, where
$\epsilon$ is the deviation from criticality, which is the same as the
{\em free-energy} singularity at a finite-$T$ transition. 
At finite-$T$, the energy and entropy each have
a stronger singularity, of the form $\epsilon^{1-\alpha}$, but with opposite
signs such that this singularity cancels in the free energy,
$F = E - T\, S$.
A analogous cancellation occurs at $T=0$, but between $E_J$ and $E_h$ since
both $E_J$ and $E_h$ have
singularities with exponent $1-\alpha$ but with amplitudes of opposite sign
such that this singularity cancels in the total energy. 
To see this note that from Eq.~(\ref{totalE})
\begin{equation}
{\partial E \over \partial h} = J {\partial E_J \over \partial h} +
h {\partial E_h \over \partial h} + E_h .
\label{eq:dedh}
\end{equation}
However, at $T=0$ where $F=E$, we have
${\partial E / \partial h} = E_h$, and so, in
this limit,
\begin{equation}
J {\partial E_J \over \partial h} +
h {\partial E_h \over \partial h} = 0 .
\label{eq:cancel}
\end{equation}
Hence, if $E_h \sim |h - h_c|^{1-\alpha}$, then $\partial E_h / \partial h$ and
$\partial E_J / \partial h$ each have singularities of the form $|h -
h_c|^{-\alpha}$, but with opposite signs such
that this singularity cancels in $\partial E / \partial h$. We have verified
that this cancellation occurs in our numerical data.
From Eqs.~(\ref{eq:dedh}) and (\ref{eq:cancel}),
$\partial E / \partial h$ has the same singularity as $E_h$, i.e.
$|h - h_c|^{1-\alpha}$, so
$E \sim |h - h_c|^{2-\alpha}$, as stated above.

We use the same set of random fields for different values of $h$ and
scale them all by a fixed overall factor. More precisely we take
$h_i = \epsilon_i h$, where the $\epsilon_i$ are chosen from a Gaussian
distribution with standard deviation {\em unity}\/,
and are the same\cite{foot1}
for all values of $h$. We use a first-order
finite difference to determine the derivative
of $E_J$ numerically and, since this is a more accurate representation of the
derivative at the midpoint of the interval
than at either endpoint, the ``specific heat'',
$C$, at $T=0$ is defined to be
\begin{equation}
C\left({h_1 + h_2 \over 2}\right) = {[E_J(h_1)]_h - [E_J(h_2)]_h \over
h_1 - h_2} ,
\label{eq:sh}
\end{equation}
where $h_1$ and $h_2$ are two ``close-by'' values of $h$, and
$[\cdots]_h$ denotes an average over
random-field configurations, which is carried
out (approximately) by
repeating the calculation for $N_{\rm samp}$ independent realizations
(samples) of
the random fields $\epsilon_i$.
We choose a sufficiently fine mesh of random-field values that the resulting
data for $C$ is smooth. Error bars are obtained by determining the specific
heat from the corresponding finite difference as in Eq.~(\ref{eq:sh})
for each sample separately, and
computing the standard deviation. The error bar is, as usual, the standard
deviation divided by $\sqrt{N_{\rm samp}-1}$.

\begin{figure}[htb]
\begin{center}
\epsfxsize=\columnwidth
\epsfbox{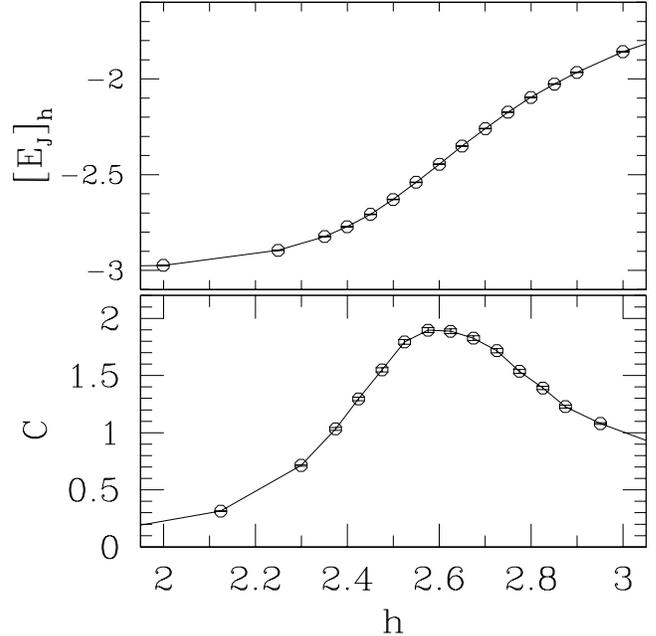}
\end{center}
\caption{The upper figure shows the average bond-energy $[E_J]_h$ per spin
as a function a function of the random-field strength $h$ for $L=16$.
The lower figure displays the resulting ``specific heat'', calculated
from Eq.~(\ref{eq:sh}) of the text.
}
\label{figCWays}
\end{figure}

In Fig. \ref{figSampleJumps} the bond energy per spin $E_J$ 
for two representative $L=8$ systems is shown
as a function of $h$. For very small
values of $h$ all spins point into the same direction and so $E_J=-3$.
For large $h$ the spins follow the random fields and so $E_J \to 0$ in this
limit.
The curves in Fig. \ref{figSampleJumps}
are stepwise constant functions because generically it is not
favorable to flip spins if the random field is increased by a small amount.
However, at certain discrete field values, the total energy of
another state, which differs in the
orientation of a cluster of spins, will become degenerate
with the energy of the ground state and for slightly
larger values of $h$ the state
with the cluster flipped will
become the new ground state. Although the {\em total}\/
energy is continuous at the
field values where the ground state configuration changes,
the bond energy, which is just
the first term in Eq.~(\ref{eq:ham}), changes discontinuously. At larger field
values the jumps in $E_J$ occur closer together and would be difficult to
distinguish on the scale of a figure. This is why we show, in
Fig.~\ref{figCWays}, data for a rather small size. Even for small sizes, the
jumps occur at different values of $h$ for different samples, and so the {\em
average}\/ value of $E_J$ is expected to be smooth.

This is illustrated 
in the upper part of Fig.~\ref{figCWays}
for $L=16$ which shows a smooth variation of $[E_J]_h$ with $h$.
The data
in the lower part of the figure
is the average specific heat, obtained 
as the numerical derivative of
the data for $[E_J]_h$ according to Eq.~(\ref{eq:sh}).
The specific heat is seen to have a peak, as
expected. We will investigate the size dependence of this peak in
Sec.~\ref{sec:results}.

\begin{figure}
\begin{center}
\epsfxsize=\columnwidth
\epsfbox{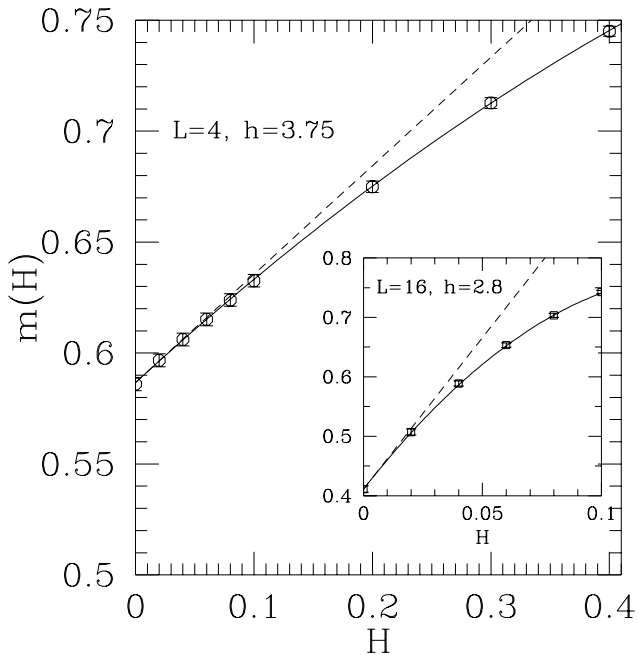}
\end{center}
\caption
{The average magnetization $m$ as a function of a uni\-form external
field $H$ near the transition for $L=4, h=3.75$ (inset: $L=16, h=2.8$).
The solid lines represent the results of fits to a parabola, while the
dashed lines display the tangents at $H=0$; i.e. their slope gives 
the susceptibility. 
}
\label{figMBhom}
\end{figure}

In addition to the specific heat, we also calculate the susceptibility by
considering the response to a small uniform external field $H$, i.e.
we consider the Hamiltonian
\begin{equation}
{\cal H}=-J\sum_{\langle i,j \rangle} S_i S_j - \sum_i h_i S_i 
- H \sum_i S_i \, .
\end{equation}
For each realization, the sign of $H$ is chosen in the
direction of the magnetization of the ground state. This prevents the
whole system from flipping when applying a magnetic field to a system
which is almost ferromagnetically ordered. The scaling behavior of the
magnetization should not be affected by this choice.
In Fig. \ref{figMBhom}, the result is shown for
system sizes $L=4$ and $L=16$ near the values of the field, where the
susceptibility attains a maximum. Near $H=0$, the data points can be fitted
very well with a parabola, the coefficient of the linear term gives
the zero field susceptibility $\chi=dm/dH|_{H=0}$. Thus, in order to
calculate the susceptibilities, we perform ground state calculations
for three different values of the uniform field $H_n=nH_L$ ($n=0,1,2$),
where, for each size, the value of $H_L$ used is
shown in Table.~\ref{tab:hvals}, along with the number of samples.
We chose the values of $H_L$ for each size as follows. For the smaller sizes
we performed several fields values, as shown in Fig.~\ref{figMBhom}, to
determine for what range of fields a parabola accurately fitted the data.
For larger sizes, finite-size scaling tells us that, near the critical point,
the characteristic field scales with $L$ as $L^{-y_H}$ where the ``magnetic
exponent'' $y_H$ is given by $(\gamma + \beta)/\nu$,
with $\gamma$ the susceptibility
exponent, and $\beta$ the order parameter exponent. As discussed further in
Sec.~\ref{sec:summary}, several calculations give $\beta \simeq 0, 
\gamma \simeq 2$, and $\nu \simeq 1.3$, and so $y_H \simeq 1.5$. We therefore
scale $H_L$ for the larger sizes by a factor of roughly $L^{-1.5}$.

For each system size,
we fit a parabola through the three data points for the {\em average}
magnetization $m(H_n)$.
To estimate
the error, we performed a jackknife analysis \cite{jackknife}
in which we divided the results for the magnetizations
(for each system size and each strength of
the disorder) into $K$ blocks, calculated the average values $K$ times,
each time omitting one of the blocks, and then performing $K$ fits. The
error bar is estimated from the variance of the $K$ results for 
the linear fitting parameter. We used $K=50$ and checked that
the result does not depend
much on the choice of $K$.
 
\begin{table}[ht]
\begin{tabular}{l|rr}
L & $N_{\rm samp}$ & $H_L$ \\\hline
4 & $10^5$ & $0.05$ \\
6 & $60000$ & $0.025$  \\
8 &  $40000$ &  $0.016$ \\
12 & $30000$ & $0.008$ \\
16 &  $23000$ &  $0.005$ \\
24 &  $27000$ & $0.0028$ \\
32 &  $15000$ &  $0.0018$ \\
48 &  $15000$ &  $9 \times 10^{-4}$ \\
64 &  $9000$ & $6 \times 10^{-4}$ \\
96 &  $3800$ &  $3\times 10^{-4}$
\end{tabular}
\caption{The maximum number of samples $N_{\rm samp}$ used, and sizes of
smallest non-zero uniform field $H_L$, for each system size $L$. As discussed
in the text, the number of samples used was larger in the vicinity of the peaks
in the susceptibility and specific heat than elsewhere. 
}
\label{tab:hvals}
\end{table}
 
\section{results}
\label{sec:results}
We have studied random-field systems with sizes from $L=4$ to $L=96$.
For each size, simulations were made for several different
values of $h$, always averaged over many realizations of the disorder.
Near the ferromagnet-paramagnet phase transition, the number of
samples used 
is the largest, ranging from $10^5$ for the smaller system
sizes to $3800$ for $L=96$ for each value $h$, as shown in
Table~\ref{tab:hvals}. 
With current algorithms,
it is in principle possible to study even larger system sizes,
such as $L=128$ or
even $L=256$,
but,
using the LEDA algorithms,
these need more memory than the $512$ MBytes available to us.
Hence we have
restricted our study to $L\le 96$, which is still much larger than sizes
that can be simulated using Monte Carlo simulations. 

In the thermodynamic limit {\em the singular part of }\/ the specific heat
diverges according to 
\begin{equation}
C_{\rm s} \approx A_{\pm} |h - h_c|^{-\alpha},
\label{ampls}
\end{equation}
where the {\em
amplitudes}\/ $A_+$ and $A_-$ refer to $h>h_c$ and $h < h_c$ respectively, and
$\alpha$ is the specific heat exponent. In addition there is a {\em regular}\/
piece of the specific heat, $C_{\rm reg}$, which is finite at the critical
point and so 
{\em dominates}\/ there if
$\alpha < 0$. In a finite system, finite-size scaling predicts
that
\begin{equation}
C_{\rm s} \sim L^{\alpha/\nu}
\widetilde{C}\left( (h - h_c) L^{1/\nu}\right)  ,
\label{eq:scale:heat}
\end{equation}
where $\nu$ is the correlation length exponent. The
specific heat peak will occur when the
argument of the scaling function $\widetilde{C}$ takes some value, $a_1$ say,
so 
the peak position $h^{*}(L)$ varies as
\begin{equation}
h^{*}(L)-h_c \approx a_1 L^{-1/\nu} ,
\label{eq:shift:max}
\end{equation}
and the value of the singular part of the specific heat at
the peak 
varies as
\begin{equation}
C_{\rm s}^{\rm max}(L) \sim  L^{\alpha/\nu} .
\end{equation}

\begin{figure}[htb]
\begin{center}
\epsfxsize=\columnwidth
\epsfbox{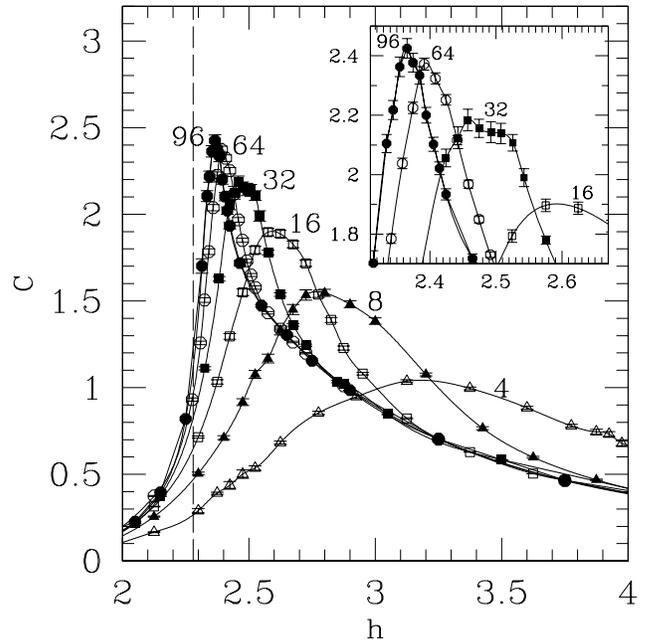}
\end{center}
\caption{``Specific heat'' $C$, calculated from Eq.~(\ref{eq:sh}),
as a function of the random-field strength $h$ 
for system sizes $L=4,8,16,32, 64$ and $96$. The vertical dashed line 
indicates the location
of the critical value of the random field, $h_c = 2.28$, see
Eq.~(\ref{eq:hc-sh}).
The inset is an enlargement of the peaks for the larger sizes.
}
\label{figHeatDElta}
\end{figure}

\begin{figure}[htb]
\begin{center}
\epsfxsize=\columnwidth
\epsfbox{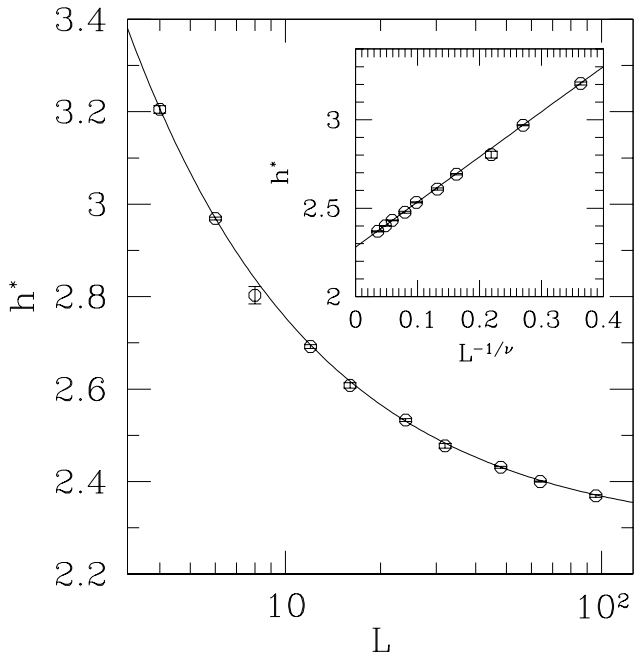}
\end{center}
\caption{A plot of the random field where the specific
heat attains its maximum, as a function of system size $L$. The
solid line shows a fit to the function 
$h^{*}(L)=h_c + a_1 L^{-1/\nu}$ with $h_c=2.28$,
$1/\nu=0.73$, and $a_1 = 2.55$.
The inset shows the data as a function of $L^{-1/\nu}$.
}
\label{figDeltaCL}
\end{figure}

In Fig. \ref{figHeatDElta} the specific heat $C$ is shown
as a function of the random-field strength $h$ for selected system sizes.
The error bars are obtained from the standard deviation of the data for
different samples, and are quite small because a large number of samples have
been averaged over, see Table~\ref{tab:hvals}.
A clear peak can be seen, which moves to the left and increases in height 
with increasing system size.
The number of samples used
is larger near the peak to compensate for the greater sample to sample
fluctuations in this region. 
For each system size, we performed  parabolic fits to the region of the peak
to obtain $h^{*}(L)$ and the height of the peak, $C^{\rm max}(L)$.
The
shift of the maximum according to Eq. (\ref{eq:shift:max}) can be used
to estimate the infinite-size critical strength 
of the random field, $h_c$ and the
correlation-length exponent $\nu$.
The best fit gives
\begin{equation}
h_c = 2.28 \pm 0.01, \qquad 1/\nu=0.73 \pm 0.02 ,
\label{eq:hc-sh}
\end{equation}
see Fig. \ref{figDeltaCL}.
We determined the probability $Q$ that the value of 
$\chi^2=\sum_{i=1}^N (\frac{y_i-f(x_i)}{\sigma_i})^2$, 
with $N$ data points ($x_i,y_i\pm\sigma_i$) fitted
to the function $f$, is worse than in the current 
fit \cite{numrec1995} to quantify the quality of the fit. Here we get
$Q=0.20$, which is fair.

\begin{figure}[htb]
\begin{center}
\epsfxsize=\columnwidth
\epsfbox{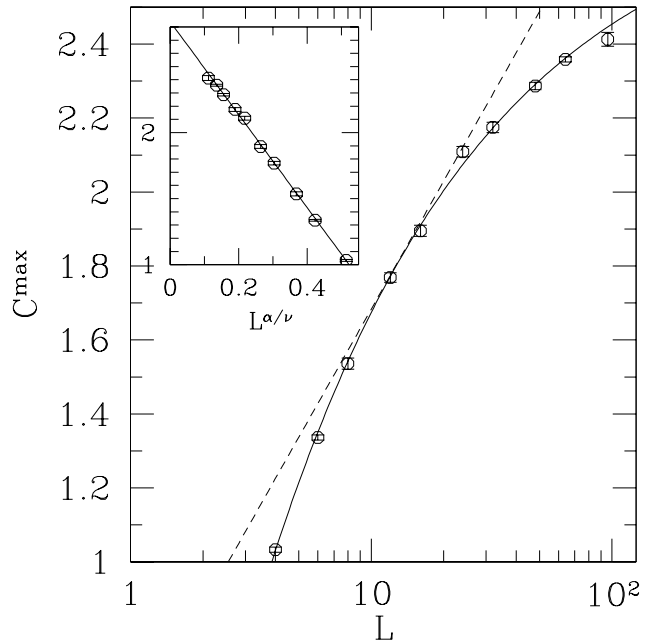}
\end{center}
\caption{The maximum $C^{\rm max}$ of the specific heat as a 
function of system 
size $L$ with logarithmically scaled $L$-axis. The dashed line
is a tangent to the data
and a comparison between it and the data
demonstrates that $C^{\rm max}$
grows  slower than 
logarithmically with system size. The solid line shows a fit to the
function $C^{\rm max}(L)=
C_{\infty}+a_2 L^{\alpha/\nu}$ with $C_{\infty}=2.84$,
$\alpha/\nu=-0.48$ and $a_2 = -3.52$.
The inset shows the data and the fit as
a function of $L^{\alpha/\nu}$.
}
\label{figHeatMaxL}
\end{figure}

Next we try to estimate the specific heat exponent by looking at how the
peak value $C^{\rm max}$ scales with $L$. If $\alpha=0$ one expects
logarithmic divergence and the simplest hypothesis is to fit the data to
\begin{equation}
C^{\rm max} = a + b \log L ,
\label{eq:logfit}
\end{equation}
where the constant term $a$ comes partly from the regular piece of the
specific heat. However, Fig.~\ref{figHeatMaxL} shows that this does not work.
A plot of $C^{\rm max}$ against $L$ (on a log scale) shows clear curvature,
suggesting that the height of the specific heat will saturate to a finite
value as $L$ increases.
If one considers
only the data points for sizes $L=4,\ldots,16$, as in Ref.
\onlinecite{rieger1995b}, 
a negative curvature is still visible, but the result is much less clear.

A peak height which saturates for $L \to \infty$ implies that $\alpha$
is negative, in which case the specific heat has a finite cusp at the critical
point, rather than a divergence. We have therefore tried a fit of the form
\begin{equation}
C^{\rm max}(L) = C_{\infty} + a_2L^{\alpha/\nu} ,
\label{eq:shmax}
\end{equation}
in which $C_{\infty}$ comes from the regular part of the specific heat,
yielding,
\begin{equation}
c_{\infty}=2.84 \pm 0.05 , \qquad
\alpha/\nu=-0.48 \pm 0.03 .
\label{eq:alpha_nu}
\end{equation}

This fit is shown in 
the inset of  Fig. \ref{figHeatMaxL}.
The quality of the fit, $Q=0.05$, is not very good. We have tried different
fits using only the larger system sizes, which increases the 
quality of the fit slightly, but the resulting error bars are very large. The
central estimate for $\alpha$ actually becomes {\em more negative}\/ if we
only include the larger sizes.
The rather poor fit may indicate difficulty in accurately estimating
the error bars for the location and height of the specific heat peak.
Our analysis suggests that the specific heat
exponent is strongly negative, in agreement with 
Rieger\cite{rieger1995b} though we cannot rule out a leading
singularity with $\alpha \simeq 0$ and a sufficiently small amplitude that it
is hard to see in our data.

To look for this possibility, 
we also tried more complicated fits including corrections to scaling of
the form
\begin{equation}
C^{\rm max}(L) = C_{\infty} + a_2 L^{\alpha/\nu} (1 + b L^{-\omega}) ,
\end{equation}
where $\omega$ is the leading correction to scaling exponent. The data did not
determine all the parameters cleanly, and the fit program\cite{fit}, which
works iteratively, converged to different results depending on the starting
values, and whether any of the parameters were held fixed. The solutions we
found were of two types: (i) the fit is the same as that in
the simpler fit of Eq.~(\ref{eq:shmax}) (i.e.
$\omega$ is essentially zero and $\alpha/\nu$ and
the other parameters are the
the same as found in the simpler fit),
(ii) $\omega$ is quite small, $ a_2$ is very large, and
$b$ is negative such that $1 + b L^{-\omega}$ is close to zero. Thus, in the
second type of fit, the data
is represented as two singularities with large amplitudes which almost cancel.
This does not seem physical. 
The fitting routine did not converge to a solution with a leading singularity
which has a small amplitude and $\alpha \simeq 0$, plus a correction term with
a larger amplitude.

We will discuss our specific heat results further in Sec.~\ref{sec:summary}. 

\begin{figure}[htb]
\begin{center}
\epsfxsize=\columnwidth
\epsfbox{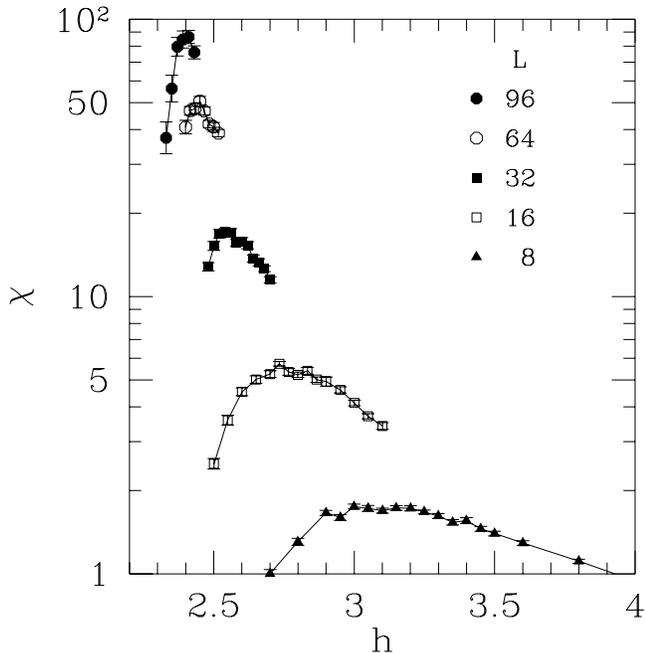}
\end{center}
\caption{Susceptibility $\chi$ as a function 
of the random-field strength $h$ 
for system sizes $L=8,16,32,64,$ and $96$. Only data near the peaks is shown
because the data away from the peaks had lower precision.
}
\label{figSuszDelta}
\end{figure}

\begin{figure}[htb]
\begin{center}
\epsfxsize=\columnwidth
\epsfbox{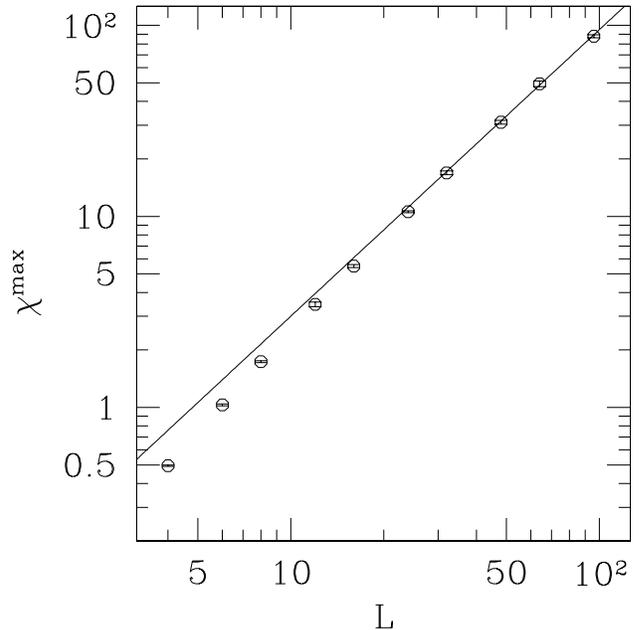}
\end{center}
\caption{The maximum $\chi^{\rm max }$ of
the susceptibility as a function of system 
size $L$ in a double logarithmic plot. 
The solid line represents a fit to the
function $\chi^{\rm max }(L)=a_3 L^{2 - \eta}$, for sizes
$L \ge 32$ yielding $2 - \eta = 1.50$ and $a_3=0.095$. }
\label{figSuszHeightL}
\end{figure}

The susceptibility $\chi$ as a function of $h$ is presented in
Fig. \ref{figSuszDelta} for selected system sizes. 
It is seen that the height of the peak grows much faster than for the
specific heat.
To analyze the divergence of $\chi$, we have again fitted parabolas
to the data points near the peak to obtain the positions
$h^{*}(L)$ and $\chi^{\rm max}(L)$
of the maximum.
By fitting the data for $L \ge 32$ to
a function $\chi^{\rm max}(L)=a_3 L^{2 - \eta}$, where $\eta$ describes the
decay of the ``connected'' correlations at criticality, we obtain ($Q=0.63$)
\begin{equation}
\eta = 0.50 \pm 0.03 , 
\label{eq:eta}
\end{equation}
see Fig. \ref{figSuszHeightL}.

Finally, we have also estimated 
$h_c$ and the correlation-length exponent from the susceptibility
data using 
Eq. (\ref{eq:shift:max}), as we did for the specific heat. 
Using only sizes $L\ge 32$ ($Q=0.84$), we find
\begin{equation}
h_c=2.29 \pm 0.01 ,
\qquad 1/\nu = 0.81 \pm 0.05 .
\label{eq:hc-chi}
\end{equation}

This estimate of $h_c$ agrees with that obtained from the specific
heat, see Eq.~(\ref{eq:hc-sh}), while the estimate for
$1/\nu$ differs from that in
Eq.~(\ref{eq:hc-sh}) by slightly more than the sum of the error bars, probably
indicating some systematic corrections to scaling. 

\section{Discussion}
\label{sec:summary}

We have determined the ``specific heat'' of the random field Ising
model at $T=0$ using optimization algorithms. The height of the peak
increases less fast than logarithmically with system size, and a finite-size
scaling analysis gives the exponents shown in Eqs.\-\ (\ref{eq:hc-sh}) and
(\ref{eq:alpha_nu}). From the analysis of the susceptibility, the
exponents shown in Eqs.\-\ (\ref{eq:eta}) and (\ref{eq:hc-chi}) are obtained.
The final results we quote are
\begin{equation}
\begin{array}{l@{\:=\:}r@{\:\pm\:}ll@{\:=\:}l}
h_c &  2.28  &  0.01, &  {\nu} & 1.32 \pm 0.07 \\
\alpha & -0.63 & 0.07, &  \eta & 0.50 \pm 0.03 .
\end{array}
\end{equation}
To determine $\nu$ and its error we have taken both the values in
Eqs.~(\ref{eq:hc-sh}) and (\ref{eq:hc-chi}) and used the difference between
them as a measure of the systematic error. The errors for $\eta$ and $h_c$
are purely statistical. The error for $\alpha$ comes both from the error in
$\nu$ and the statistical error in $\alpha/\nu$.

Our results for $h_c$ are compatible with the values
$2.29\pm 0.04$ \onlinecite{alex-rfim3},
$2.26 \pm 0.01$ \onlinecite{angles-d-auriac1997},
and $2.270\pm 0.005$ \onlinecite{mf01} obtained from
ground-state calculations of systems of similar size.
Values of $\nu$ obtained from ground-state calculations
are $1.37 \pm 0.09$ \onlinecite{mf01} and $1.19 \pm 0.08$,
\onlinecite{alex-rfim3}, which agree well with our result.
Ref. \onlinecite{angles-d-auriac1997} argued for a first-order
transition, but assuming scaling with respect to the
field, a value of $\nu=1.25 \pm 0.06$ was estimated, also in agreement with our
result. However, if
a power law correction to scaling was taken into account, instead the result
$1.52$ (without error bars) was found.

The scaling exponent $\eta$ describing the susceptibility, has not been
obtained from exact ground-state calculations so far. In a Monte-Carlo
simulation\cite{rieger1995b} a value of $0.50\pm 0.05$ was found, which
is compatible with our result.


The most significant result of this paper is that for the specific heat,
namely $\alpha=-0.63(7)$. This agrees well with the values
$\alpha/\nu = -0.45 \pm 0.05, \nu = 1.1 \pm 0.2$ found by
Ref.~\onlinecite{rieger1995b} and
$\alpha = -0.55 \pm 0.20$ found by Ref.~\onlinecite{nowak1998}, both
using Monte Carlo simulations on small systems.
However, as we shall now see, it appears
inconsistent with values for other exponents and expected scaling relations.

At conventional second-order phase transitions, all exponents can be related to
{\em two}\/
(e.g. $\nu$ and $\eta$) by scaling relations. However, because the fixed
point of the RFIM is at $T=0$ with temperature a ``dangerous irrele\-vant
variable'', a modified set of scaling relations has been
proposed\cite{bm85,fisher85,grinstein,villain}, which involve {\em three}\/
independent exponents. Scaling relations which do not involve the space
dimension, e.g. 
\begin{equation}
\alpha + 2\beta + \gamma = 2 , 
\label{eq:rushbrooke}
\end{equation}
are unchanged, but ``hyperscaling'' relations involving the space dimension
$d$, have $d$
replaced by $d-\theta$, where $\theta$, the third exponent, is the scaling
exponent for the temperature at the fixed point. An example of a hyperscaling
relation which is relevant to the specific heat is
\begin{equation}
(d - \theta) \nu = 2 - \alpha .
\label{eq:sh-hyper}
\end{equation}
Gofman et al.\cite{gaahs} have proposed that
the Schwartz-Soffer\cite{ss} inequality, which can be expressed as
$\eta \ge 2 - \theta$, is an equality, in which case there are only two
independent exponents again (though the hyperscaling relations are different
from those in conventional two-exponent scaling). Our results are consistent
with this, since $\beta \simeq 0$ implies that $\theta \simeq 1.5$, see e.g.
Ref.~\onlinecite{mf01}, and we have already found that $\eta$ is about 0.50,
see Eq.~(\ref{eq:eta}).

Other works have found\cite{rieger1995b,alex-rfim3} 
$\beta \simeq 0$ (the most accurate value is $0.017 \pm 0.005$ in
Ref.~\onlinecite{mf01}), and our value for $\gamma$, obtained from
$\gamma \equiv (2 - \eta)\nu$ is about $2.0$ in agreement with series
expansion work of Gofman et al.\cite{gaahs}.
Hence Eq.~(\ref{eq:rushbrooke}) predicts $\alpha \simeq
0$, quite different from the value of about $-0.63$ that we find by direct
calculation.

As noted above,
the result
$\beta \simeq 0$ implies that $\theta \simeq 1.5$,
so Eq.~(\ref{eq:sh-hyper}) gives $\alpha \simeq 2 - 1.5 \nu$.
Using our value of $\nu = 1.32 \pm 0.07$ this yields 
$\alpha = 0.0 \pm 0.15$. In other words, Eq.~(\ref{eq:sh-hyper}) also
predicts that $\alpha$ is close to zero.

We have seen that the two scaling relations above
would be consistent if we inserted
$\alpha \simeq 0$, which is the experimental value\cite{belanger}.
However,
by direct calculation, we obtain a strongly negative result, $\alpha \simeq
-0.63$, consistent with earlier work\cite{rieger1995b} on much smaller sizes.
Thus the problem with the value of the
specific-heat exponent has now been strongly reinforced by
our calculations on much larger lattices.

Possible explanations for this discrepancy are:
\begin{itemize}
\item
The specific heat diverges but slower than logarithmically. Examples of this,
which are known to occur in other systems, are a fractional power of a log and
a log-log variation. However, there are no calculations which predict this
type of behavior for the RFIM. Furthermore, attempts to fit our data to this
type of behavior were not very successful. A related possibility, which does
not seem impossible looking at Fig.~\ref{figHeatDElta}, is that
$\alpha = 0$ might be realized by a {\em jump}\/ in the specific heat, with 
a lower value in the ferromagnetic region, the opposite of what occurs in mean
field theory. 
\item
The regular contribution to the specific heat varies rapidly near the critical
point. Since $\beta \simeq 0$ the magnetization increases very rapidly below
$h_c$ (leading to the very rapid drop in the specific heat seen in
Fig.~\ref{figHeatDElta}). If much of this drop comes from the
regular part of the specific heat it would be
difficult to extract the singular part.
\item
There are very strong singular corrections to finite-size sca\-ling which
leads to 
the most singular term in the specific heat being numerically
small compared with
correction terms, even for the quite large range of sizes that we have studied
here.
If there {\em are}\/
strong corrections to
scaling, perhaps 
the values of other exponents, in addition to $\alpha$, could be affected too.
\item
Scaling does not hold.
We find this possibility to be the least palatable.
\end{itemize}

Since
$\beta
\simeq 0$, it is interesting to ask whether the transition might be first
order and whether this might be the
origin of the surprising value of $\alpha$.
The transition at low-$T$ {\em is}\/ first order in mean field theory for
field distributions with a minimum at zero field\cite{aharony}. A first order
transition for Gaussian distribution has also been suggested for dimension
less than four based on series expansion work\cite{houghton}.
If the transition is first order,
it must be very weakly so, since fluctuation effects are very large.
Furthermore, one would then expect a latent heat, which, in a finite-size
system, gives a specific heat diverging as the volume $L^d$. In our
results, we do not
see {\em any}\/ divergence, let alone a strong one like this. In addition, the
most detailed numerical study\cite{mf01} claims that $\beta$ while very small,
is {\em greater}\/ than zero. Even if the transition were ultimately
first order,
the effective exponents found should be those of the close-by second order
transition, and so should satisfy scaling.  We therefore don't feel that the
possibility of a first order transition explains why our value for $\alpha$ does
not satisfy scaling. 

In addition to critical exponents, it is useful to discuss amplitude ratios,
since these are also universal, see Ref.~\onlinecite{pha}
and references therein.
For the specific heat amplitudes,
$A_+$ and $A_-$, defined in Eq.~(\ref{ampls}), one can show\cite{ratio1}
that $A_+/A_- = 1$ for a logarithmic divergence ($\alpha = 0$).
Furthermore, for $n$-component models without random fields
one has\cite{pha,bervillier} $A_+/A_- >1$
for $\alpha < 0$ and $A_+/A_- <1$ if $\alpha > 0$. This implies that, for both
signs of $\alpha$, the
specific heat decreases from its peak faster on the paramagnetic
side than on the ferromagnetic side (we are grateful to D.~Belanger for
pointing this out). By contrast, the situation is
reversed in our data, see Fig.~\ref{figHeatDElta} where the specific heat
appears to decrease
faster for $h < h_c$. Whether this indicates that the amplitude
ratio is very different in the presence of random fields,
or that corrections to scaling are large compared with the leading singularity
for this range of sizes remains to be seen.

Clearly more work is needed to understand the specific heat of the RFIM. Since
several recent large-scale numerical calculations, including ours, 
have used fairly sophisticated algorithms, it is unlikely
that a numerical breakthrough is imminent. Hence a better theoretical
understanding, especially of corrections to scaling, will be needed to sort
out this problem.

{\em Additional Note:}\/ After this work was submitted we received the final
version\cite{mf01final} of Ref.~\onlinecite{mf01} in which, motivated by our
work, they computed the bond energy using ground state methods. They
did not numerically differentiate the data to get the specific heat but
directly analyzed data for the bond energy at the bulk critical field, the
dashed line in Fig.~\ref{figHeatDElta}.  The size dependence involves the
exponent $(1-\alpha)/\nu$ from which they find results compatible with $\alpha
= 0.$ That they get a different result from ours by, in effect, considering a
different region of the scaling function in Eq.~(\ref{eq:scale:heat}),
indicates that there are large corrections to finite size scaling even for
such large sizes, or possibly that $\alpha \simeq 0$ corresponds to a
discontiniuty in the specific heat. Both these possibilities were discussed
above.  Further work is needed to clarify the situation.

\acknowledgements
We thank D.~P.~Belanger for stimulating dis\-cussions and 
Alan Middleton for giving helpful hints, showing us an advance copy of
Ref.~\onlinecite{mf01}, and commenting on an earlier version of this paper.
The simulations were performed at the Paderborn Center
for Parallel Computing in Germany and on a workstation
cluster at the Institut f\"ur Theoretische Physik, University of G\"ottingen,
Germany. AKH acknowledges financial support from the DFG (Deutsche 
Forschungsgemeinschaft)
under grant Ha 3169/1-1. APY acknowledges support
from the NSF through grant DMR 0086287.


\begin{thebibliography}{99}
\bibitem{foot1}
    Actually we used more samples for some values of $h$ than for others.
    Hence, for each data point, we only used realizations where the ground
    states were calculated at both the left and the right boundary of the
    corresponding interval.
\end{thebibliography}
\end{document}